# Numerical Study of Liquid Flow and Heat Transfer in Rectangular Microchannel with Longitudinal Vortex Generators

Amin Ebrahimi[1], Ehsan Roohi[1, *], Saeid Kheradmand[2]

1- High-Performance Computing (HPC) Laboratory, Department of Mechanical Engineering, Faculty of Engineering, Ferdowsi University of Mashhad, Mashhad, P.O. Box 91775-1111, Khorasan Razavi, Iran.

2- Department of Mechanical and Aerospace Engineering, Malek-Ashtar University of Technology, Shahin-shahr, P.O. Box 83145/115, Isfahan, Iran.

\* - Corresponding Author, Email: e.roohi@um.ac.ir, Phone: +989155153124, Address: High Performance Computing (HPC) Laboratory, Department of Mechanical Engineering, Faculty of Engineering, Ferdowsi University of Mashhad, Mashhad, P.O. Box 91775-1111, Khorasan-Razavi, Iran.

**Abstract**

The liquid flow and conjugated heat transfer performance of single-phase laminar flow in rectangular microchannels equipped with longitudinal vortex generators (LVGs) are numerically investigated. Deionized-water with temperature-dependent thermo-physical properties is employed to conduct the simulations. Three dimensional simulations are performed using an open-source flow solver based on finite volume approach and SIMPLEC algorithm. Five different configurations of the microchannel with different angles of attack of the LVGs are considered. Simulation results are compared with available experimental data and a deviation below 10% is achieved. The results show that there is a 2-25% increase in the Nusselt number for microchannels with LVGs, while the friction factor increased by 4-30%, for Reynolds number ranged from 100 to 1100. Except one at Re=100, the overall performance of the all configurations of microchannels with LVGs is higher than one.

# 1. Introduction

Studying fluid flow and heat transfer in Micro/Nano systems have become one of modern and active research areas and have attracted many researchers in few past decades. Among microfluidic systems, microchannels, due to their high area-to-volume ratio and less amount of coolant requirement, play an important role and are widely employed in cooling of electronic devices, medical industries, chemical engineering, automotive heat exchangers, laser equipment and aerospace technology.

After the pioneering work of Tuckerman and Pease in 1981 [1], where the concept of microchannel heat sink was first proposed, numerous investigations have been focused on study of fluid flow and heat transfer characteristics in microchannels, as reviewed by Morini [2], Agostani et al. [3] and Adham et al. [4]. Generally, due to higher heat transfer coefficients of liquid coolants compared to gaseous coolants, liquid coolants are utilized with microchannel heat sinks [5]. Harms et al. [6] found critical Reynolds number of 1500, by investigating forced convection of developing laminar water flow in an approximately 1000μm deep rectangular microchannel. Li et al. [7] numerically studied the heat transfer process in two non-circular microchannels under laminar flow condition for Reynolds numbers lower than 500. They reported that for liquid flow in microchannels sized to a hydraulic diameter of tens of micrometer, the conventional Navier-Stokes and energy equations with no-slip boundary condition based on the continuum assumption are still valid and could adequately predict fluid flow and heat transfer characteristics.

The methods of heat transfer augmentation can be classified in main flow or secondary flow enhancement groups as well as active (needs external power) or passive (requires no external power) methods [8-10]. Johnson and Joubert studied the impact of vortex generators (VGs) on heat transfer performance for the first time in 1969 [11]. VGs can be used in forms of wings, winglets, fins, ribs, inclined blocks and protrusions [12]. The pressure difference between two sides of a vortex generator make the flow separates from side edges and generates transverse, longitudinal and horseshoe vortices [13-15]. Generated vortices are mainly longitudinal when the angle of attack of VGs is small and they are mainly transverse when VGs are perpendicular to the main flow direction [16, 17]. Generated vortices by VGs can effectively disturb the flow boundary-layer, distort the temperature field in channel, form secondary flow and transporting central flow next to the wall and vice versa [12]. Experimental studies of Fiebig et al. [18] indicated that local heat transfer in laminar regime for channels equipped with VGs could be boosted by a factor of three relative to one without VGs. Wu and Tao [19] experimentally and numerically studied the effect of various winglet attack angels on heat transfer performance and they found that the average Nusselt number of surfaces enhances by increasing the attack angle. However, longitudinal vortices illustrate better heat transfer performance and lower flow loss compared to transverse vortices [9, 16]. Recently, the effect of using nanofluids in Macro-channels with VGs has attracted more attention [12, 20-22].

Liu et al. [23] employed winglet type VGs with different number of pairs and angles of attack in microchannels to enhance heat transfer for Reynolds numbers ranging between 170 and 1200. They found a significant heat transfer enhancement with larger pressure drop. They also reported that the range of critical Reynolds number decreases by adding LVGs compared to one without LVGs. Chen et al. [24] extended the experiments of Liu et al. [23] for different configurations with different hydraulic diameter and height of the microchannel having LVGs. Lan et al. [25] studied flow characteristics and heat transfer in a rectangular microchannel with dimple/protrusions numerically at Reynolds numbers ranging between 100 and 900. Furthermore, their results show that using a dimple/protrusion technique in microchannel could enhance heat transfer with higher pressure drop. Mirzaee et al. [26] proposed a novel mechanism for enhancing heat transfer in microchannels using elastic VGs. They performed two dimensional numerical simulations in a range of the Reynolds number varies from 100 to 500 with water as the working fluid. They found 15-35% increase in the average Nusselt number as well as a 10-70% increase in the friction factor. Also, they reported that the elastic VGs give a higher Colburn/friction factor ratio in comparison to the rigid VG for all Reynolds numbers studied in their research. Recently, Hsiao et al. [27] numerically investigated the application of LVGs in microchannels to enhance fluid mixing.

By reviewing the available literatures published, the use of LVGs in microchannels for cooling electronic devices is still emerging and more studies are essential to achieve deeper understanding of heat transfer and flow features of these devices. It can be noted that a systematic and detailed analysis for the heat transfer and flow characteristics through microchannels with LVGs is not available. Therefore, due to promising performance of this state-of-the-art technique for heat transfer enhancement, three-dimensional simulations are implemented to study conjugated heat transfer characteristics and liquid flow behavior in rectangular microchannels with LVGs using finite volume approach. The numerical simulations are performed to obtain the local variation in flow parameters for liquid flow through microchannels with LVGs. Effects of different configurations and the Reynolds number on the heat transfer and flow characteristics in laminar regime are investigated in this paper. According to the author's knowledge, no numerical study is performed to peruse the effects of using LVGs in microchannels.

## 2. Model Descriptions

### 2.1. Geometric configurations and computational domain

Three dimensional simulations are carried out on five different configurations of the microchannel. The microchannels are equipped with two pairs of winglet type VGs. Figure 1 illustrates the physical model and relevant parameters of designed microchannels. The length of the microchannel (L) and the width (W) equals 150H and 20H, respectively, where H is the height of the microchannel. The height of winglets is equal to microchannel height and their chord length (l) and thickness (b) are 5H and H/4, respectively. The flow is described in a Cartesian coordinate system in which x, y and z are span wise, normal and stream-wise coordinates, respectively. Additionally, y stands for the winglet pitch direction. Table 1 presents the details of the geometric parameters used to locate LVGs in the microchannel for different configurations and their schematic diagrams. The designed microchannels can be constructed by micro-fabrication technology such as standard etching processes [23, 24, 28]. Due to a symmetric arrangement of the microchannels and winglets, the computations are done on the hatched region in figure 1.

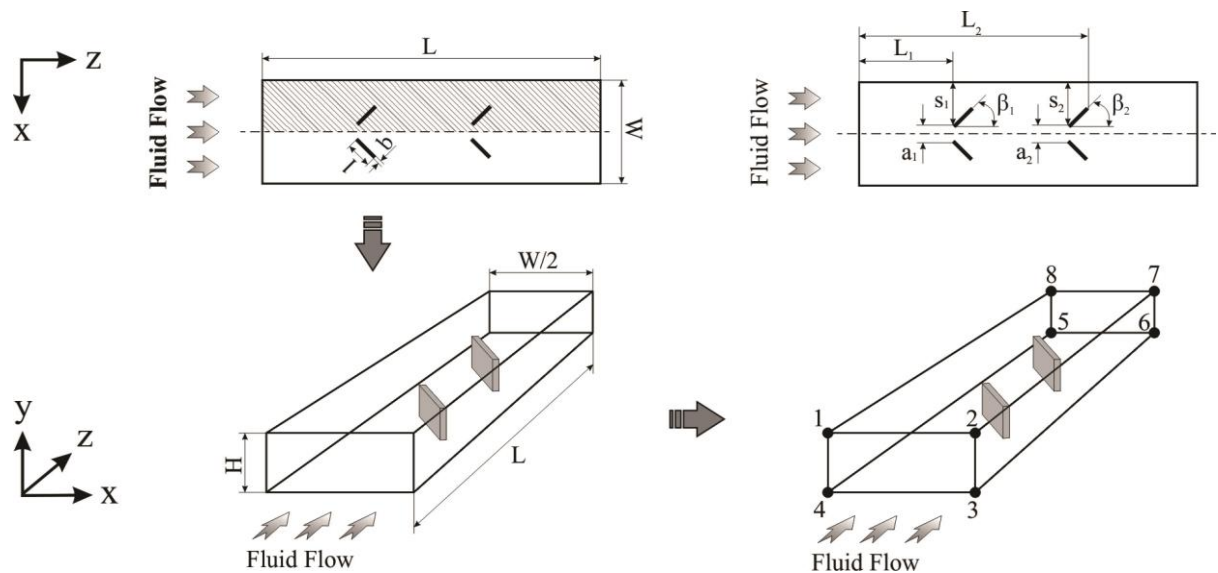

Figure 1- Physical model and relevant geometrical parameters.

Table 1- Schematic diagrams of the studied microchannels and corresponding parameters for positioning LVGs in the microchannel.

| Microchannel | Schematic diagram | $L_1$ | $L_2$ | $a_1$ | $a_2$ | $s_1$ | $s_2$ | $\beta_1, \beta_2$ |
|---|---|---|---|---|---|---|---|---|
| M0 | 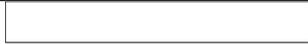 | - | - | - | - | - | - | - |
| M1 | 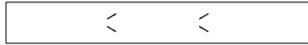 | 50H | 100H | 4H | 4H | 8H | 8H | 30°, 30° |
| M2 | 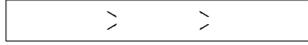 | 50H | 100H | 9H | 9H | 5.5H | 5.5H | 150°, 150° |
| M3 | 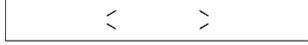 | 50H | 100H | 4H | 9H | 8H | 5.5H | 30°, 150° |
| M4 | 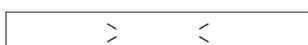 | 50H | 100H | 9H | 4H | 5.5H | 8H | 150°, 30° |

2.2. Mathematical models, governing equations and boundary conditions

Microchannels and LVGs have been simulated as a conductive material like silicon and the effects of surface roughness are deemed to be neglected [23, 24]. Deionized-water with temperature-dependent thermo-physical properties is used as the coolant and modeled to be Newtonian. Ferrouillat et al. [29] reported that longitudinal vortices are generated on a quasi-steady phenomena. The present model is derived based on continuum flow approach and steady-state conservation equations of mass, momentum and energy. The deionized-water flow is assumed incompressible and due to the low inlet velocity and the small winglet pitch the flow is assumed to be laminar. Moreover, radiation and gravity effects are ignored in this study. The resulting governing equation can be written as follow:

$$\frac{\partial(u\varphi)}{\partial x}+\frac{\partial(v\varphi)}{\partial y}+\frac{\partial(w\varphi)}{\partial z}=\frac{\partial}{\partial x}(\Lambda_\varphi \frac{\partial \varphi}{\partial x})+\frac{\partial}{\partial y}(\Lambda_\varphi \frac{\partial \varphi}{\partial y})+\frac{\partial}{\partial z}(\Lambda_\varphi \frac{\partial \varphi}{\partial z})+S_\varphi \qquad (1)$$

As summarized in table 2, $\varphi$ is corresponding to the dependent parameters $u$, $v$, $w$ and $T$. Additionally, $\Lambda_\varphi$ and $S_\varphi$ stand for the diffusion coefficient and source terms, respectively. The thermo-physical properties of deionized-water and silicon are presented in table 3. In table 3, $\mu$ is dynamic viscosity, $\rho$ is density, $c_p$ is specific heat capacity, $k$ is thermal conductivity, and $p$ is static pressure. The required boundary conditions at the inlet, outlet, and the symmetry plane as well as the solid-liquid interface, according to figure 1, are introduced as follow:

Inlet boundary (1-2-3-4): $u = v = 0, \quad w = U_{in}, \quad T = T_{in} = 298.15 \, (K)$

Outlet boundary (5-6-7-8): $\dfrac{\partial u}{\partial z} = \dfrac{\partial v}{\partial z} = \dfrac{\partial w}{\partial z} = 0, \quad \dfrac{\partial T}{\partial z} = 0$

Symmetry plane (2-3-6-7): $\dfrac{\partial v}{\partial x} = \dfrac{\partial w}{\partial x} = 0, \quad \dfrac{\partial T}{\partial x} = \dfrac{\partial T_{sl}}{\partial x} = 0, \quad u = 0$

Heated wall (1-2-7-8): $u = v = w = 0, \quad T = T_{sl} = T_{wall} = 323.15 \, (K)$

Adiabatic walls (3-4-5-6 and 1-4-5-8): $u = v = w = 0, \quad \dfrac{\partial T}{\partial y} = \dfrac{\partial T_{sl}}{\partial y} = 0$

Solid-liquid interfaces (LVGs surfaces): $u = v = w = 0, \quad k\dfrac{\partial T}{\partial \mathbf{n}} = k_{sl}\dfrac{\partial T_{sl}}{\partial \mathbf{n}}$

Where $\mathbf{n}$ is a normal vector on LVGs drawn outward the boundary. The subscript *sl* represents solid characteristics.

Table 2- List of governing equations.

| Equations | $\varphi$ | $\Lambda_\varphi$ | $S_\varphi$ |
|---|---|---|---|
| Continuity | 1 | 0 | 0 |
| x-Momentum | $u$ | $\dfrac{\mu}{\rho}$ | $\dfrac{-1}{\rho}\dfrac{\partial p}{\partial x}$ |
| y-Momentum | $v$ | $\dfrac{\mu}{\rho}$ | $\dfrac{-1}{\rho}\dfrac{\partial p}{\partial y}$ |
| z-Momentum | $w$ | $\dfrac{\mu}{\rho}$ | $\dfrac{-1}{\rho}\dfrac{\partial p}{\partial z}$ |
| Energy (Fluid zone) | $T$ | $\dfrac{k}{(\rho c_p)_{fl}}$ | 0 |
| Energy (Solid zone) | $T_{sl}$ | $k_{sl}$ | 0 |

Table 3- Temperature-dependent thermo-physical properties of deionized-water and silicon.

|  | Silicon[30] | Deionized-water[31, 32] |
| --- | --- | --- |
| μ(Pa.s) |  | $0.0194 - 1.065 \times 10^{-4} T + 1.489 \times 10^{-7} T^2$ |
| k(W/m K) | $290 - 0.4T$ | $-0.829 + 0.0079T - 1.04 \times 10^{-5} T^2$ |
| $c_p$(J/kg K) | $390 + 0.9T$ | $5348 - 7.42T + 1.17 \times 10^{-2} T^2$ |
| ρ(kg/m³) | 2330 | 998.2 |

2.3. Numerical procedures and parameter definitions

In the present study, OpenFOAM which is an open-source package for computational fluid dynamics (OpenFOAM documentation, 2014) is used. Structured non-uniform grids are utilized to discretize the computational domain in which by using an expansion ratio for cells width adjacent to the winglets and channel walls, cells are appropriately fine in order to improve accuracy. The conservation equations are discretized using finite volume method based on SIMPLEC algorithm. The governing equations are solved iteratively until the convergence criterion satisfied which is defined when the scaled residuals of the momentum, continuity and energy equations attain a value less than $1.0 \times 10^{-5}$, $1.0 \times 10^{-5}$ and $1.0 \times 10^{-7}$, respectively. To present the results of numerical simulations the subsequent parameters are defined. The Reynolds number (*Re*) based on the hydraulic diameter ($D_h$) of microchannel is defined as follow.

$$Re = \frac{U_{in} D_h}{\nu} \quad (2)$$

where

$$D_h = \frac{4 A_{ch}}{P_w} = \frac{2WH}{W + H} \quad (3)$$

Where, $A_{ch}$ is cross sectional area of the inlet and $P_w$ is wetted perimeter of the microchannel. Additionally, $U_{in}$ is the inlet velocity and $\nu$ is the kinematic viscosity at the inlet. The Nusselt number (*Nu*) and the mean Nusselt number ($Nu_m$) can be calculated by the following relations.

$$Nu = \frac{hD_h}{k} \quad (4)$$

$$Nu_m = -\frac{D_h}{k_{fl,m}} \ln\left(\frac{(T_{wall}-T_{in})}{(T_{wall}-T_{out})}\right) \frac{\dot{m}c_{p,m}}{A_{ht}} \quad (5)$$

where

$$h = \frac{Q}{A_{ht}\Delta T} \quad (6)$$

$$Q = \dot{m}c_p(T_{out}-T_{in}) \quad (7)$$

$$\Delta T = \frac{(T_{wall}-T_{in})-(T_{wall}-T_{out})}{\ln[(T_{wall}-T_{in})-(T_{wall}-T_{out})]} \quad (8)$$

In the above relations $Q$ is the total heat rate, $A_{ht}$ is the area of heated wall, $\dot{m}$ is mass flow rate of deionized-water, $c_p$ is specific heat capacity of deionized-water and $T_{in}$ and $T_{out}$ are bulk fluid temperatures at the inlet and the outlet, respectively. In Eq. 5, $k_{fl,m}$ and $c_{p,m}$ denote thermal conductivity and specific heat capacity of fluid at the arithmetic mean temperature of the inlet and the outlet $(T_{in}+T_{out})/2$, respectively. The apparent friction factor ($f$) and required pumping power ($P_p$) can be calculated by subsequent relations:

$$f = \frac{2\Delta p}{\rho_f U_{in}^2}\frac{D_h}{L} \quad (9)$$

$$P_p = \Delta p \dot{V} \quad (10)$$

Where

$$\Delta p = (\bar{p}_{out} - \bar{p}_{in}) \quad (11)$$

$$\bar{p} = \frac{\int p \, dA}{\int dA} \quad (12)$$

In these equations, $\rho_f$ is fluid density, $\Delta p$ is pressure drop across the computational domain and $\bar{p}_z$ is area-weighted static pressure at the cross section which pressure coefficient is being evaluated in stream-wise direction. Moreover, $\dot{V}$ is volumetric mass flow rate of deionized-water in the microchannel.

## 3. Grid independency and model verification

Six grids with different sizes are employed for the grid independence test on M1 configuration. The grid used have sizes of 276,000 (very coarse), 381,000 (coarse), 502,000 (intermediate), 624,000 (fine), 652,000 (very fine) and 730,000 (extremely fine). The results of the $Nu_m$ for different grid sizes at $Re=600$ is presented in table 4. It is noticed that there is no difference in the computed results between the very fine and the extreme fine grids. Hence, by a compromise between required accuracy, time usage and computation costs, the very fine grid is selected for the simulations.

Table 4- Comparison of the mean Nusselt number for different grid sizes at $Re=600$.

| Number of cells | Predicted $Nu_m$ | Percentage difference of $Nu_m$ |
| --- | --- | --- |
| 276,000 (very coarse) | 8.4693 | - |
| 381,000 (coarse) | 8.3143 | -1.863% |
| 502,000 (intermediate) | 8.2748 | -0.478% |
| 624,000 (fine) | 8.2276 | -0.573% |
| 652,000 (very fine) | 8.2162 | -0.139% |
| 730,000 (extremely fine) | 8.2162 | 0.000% |

In order to verify the ability of the solver to predict accurate and reliable results, validation of the code was performed against the experimental results presented by Liu et al. [23]. The experimental equipment in their study consists of a rectangular microchannel made of silicon with a hydraulic diameter of 187.5µm and two centimeters long. They used deionized-water as the coolant. The Reynolds number based on the channel hydraulic diameter varies from 170 to 1200. Computations are done on only one symmetrical part of two different configurations presented in [23] with the same boundary conditions; One a smooth channel and the other one a microchannel equipped with three pairs of winglets in which the angle of attack of the winglets are 30, 150 and 30 degrees, respectively. More detailed information about the test cases and experiments setup can be found in [23]. The effects of both constant and variable thermo-physical properties on the results are considered.

Figure 2 indicates the results of the predicted $Nu_m$ compared to the experimental results. A good agreement between numerical results and the experimental results of Liu et al. [23] is achieved. The maximum deviation of numerical results from experimental results for smooth microchannel and microchannel with LVGs is less than 11% and 6%, respectively. It is worth mentioning that the average uncertainties caused by the experimental apparatuses in determining the $Nu_m$ for the smooth microchannel and the microchannel with LVGs is 19.6% and 19.07%, respectively [23]. Furthermore, more deviation is observed by applying constant thermo-physical properties.

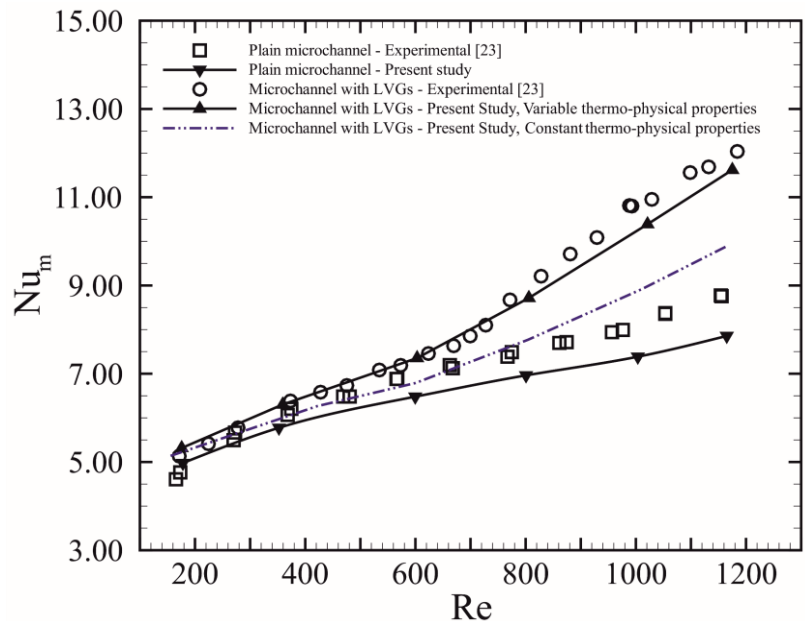

Figure 2- Comparison of the present numerical results with experimental data of Liu et al. [23] for various Reynolds numbers.

## 4. Results and discussions

Figure 3 demonstrates the non-dimensional velocity magnitude ($U/U_{in}$) contours and the streamlines on a plane located at half of the microchannel height ($y=0.5H$) for M1 configuration at $Re=300$, 700 and 1100. It is observed that high velocity zone forms between LVGs and nearby the side walls of the microchannel. It can be seen that the recirculation regions behind the LVGs increase with increasing the Reynolds number. Furthermore, it is clear that fluid mixing intensifies by increasing the Reynolds number. It is found that interaction between vortices occurs due to the generation of the shear layer behind the LVGs. At higher Reynolds numbers the flow covers longer distances in microchannels with LVGs in relation to smooth microchannel because the fluid passes through a wavy path downstream of the LVGs. It can be explained that the generated vortices due to the presence of LVGs become stronger with increasing the Reynolds number and hence bring more heat transfer enhancement in respect to smooth microchannel. Additionally, this phenomenon may be attributed to transition from laminar to turbulent condition. Similar inference was also reached by Ma et al. [33] and Liu et al. [23]. The other configurations with LVGs represent similar behaviors.

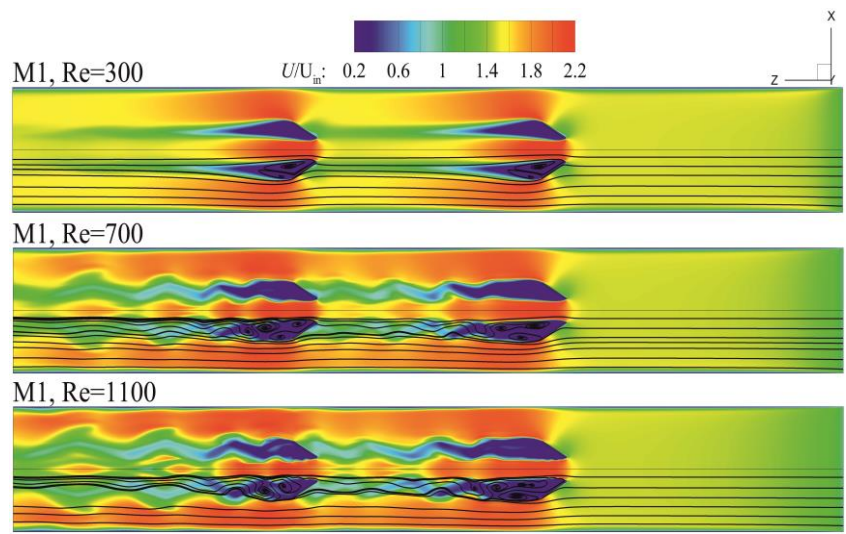

Figure 3- Non-dimensional velocity magnitude ($U/U_{in}$) contours and the streamlines for M1 configuration at $Re$=300, 700 and 1100. ($y$=0.5H)

The non-dimensional temperature contours in M1 microchannel and also on the surfaces of the LVGs are presented in figure 4 for various Reynolds numbers. It is observed that the temperature distribution inside the solid zone is not uniform and its average temperature is high. The LVGs of the first row has a higher average temperature relative to that of the LVGs in the second row. It is seen that the interactions between coolant and hot surfaces of the top plate and LVGs change the temperature distribution. Increasing the Reynolds number causes more changes in temperature distribution and decrease the thermal boundary-layer thickness on the microchannel walls and the windward side of LVGs. Presence of longitudinal and horseshoe vortices and decreasing the thermal boundary-layer thickness results higher temperature gradients and therefore higher heat transfer rate and lower average temperature of LVGs. There is a higher temperature difference between the coolant and hot surfaces at higher

Reynolds numbers which results higher heat transfer rate. This can be described that at higher Reynolds numbers forced convection dominates diffusive heat transfer, and vice versa [34-36]. Additionally, higher bulk and local fluid temperature are seen in microchannels with LVGs which results lower fluid viscosity. This can make the fluid, less stable, augments eddy generation and fluid mixing and destabilization which consequences more heat transfer enhancement. It can be concluded that fluids with lower viscosity will perform better thermal performance in microchannels with LVGs. The heat transfer behavior is similar for the other configurations with LVGs. A higher heat transfer area in microchannels with LVGs compared to smooth microchannels also is another reason for larger heat transfer enhancement capability.

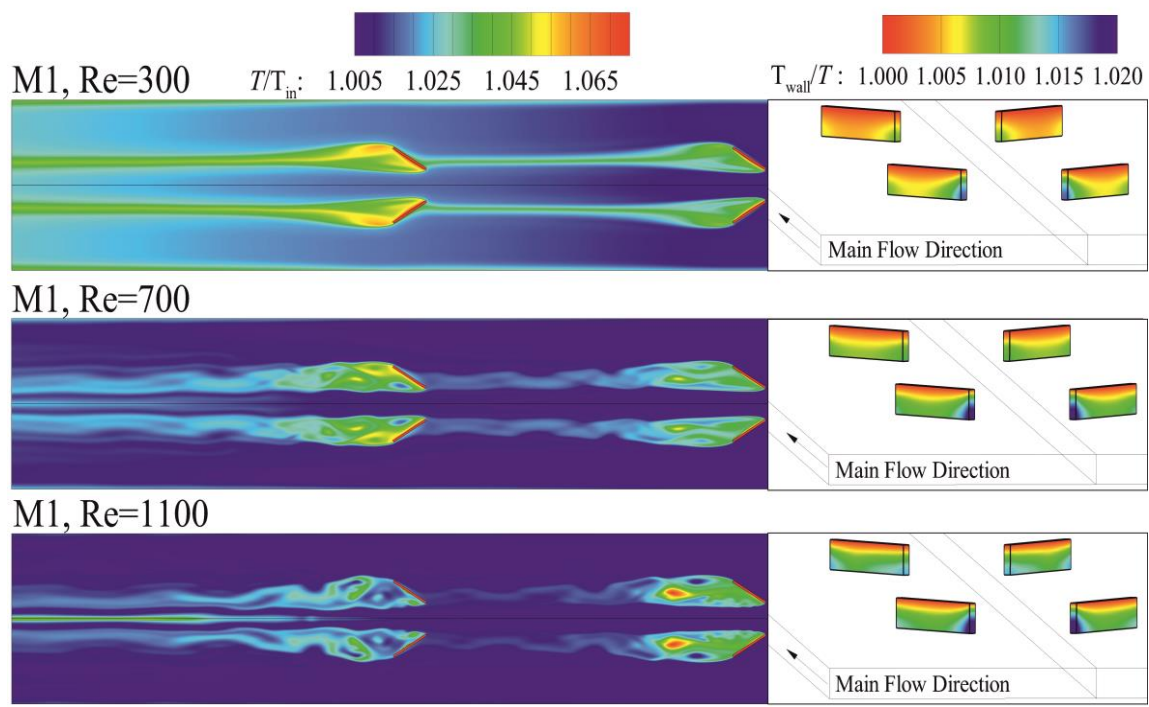

Figure 4- Non-dimensional temperature contours in M1 microchannel and also on the surfaces of the LVGs at $Re$=300, 700 and 1100. ($y$=0.5H)

Figure 5 shows the contours of non-dimensional temperature ($T/T_{in}$) on different cross sections along the stream-wise direction for different configurations at $Re$=700. For all the configurations, two high temperature zones are seen behind the LVGs due to the presence of horseshoe vortices along the lateral side of the LVGs. The two high temperature zones merge together as the flow moves downstream and spreads gradually to a region adjacent to the heated wall while the vortices lose their strength. It is found that longitudinal vortices are incapable to form in full scale due to high aspect ratio (W/H) of the microchannel cross section for all the configurations. It is seen that an up-wash and a down-wash flow is induced in proximity of the LVGs, indicating that intense secondary flow and strong fluid mixing are produced by the LVGs. The down-wash flow brings the hotter fluid adjacent to the heated wall to the central region and thus makes the thermal boundary-layer thickened. The up-wash flow, simultaneously, transfers the cooler fluid from the central region to the heated wall and makes the boundary-layer thinner. So, as the temperature gradient increases near the walls, the heat transfer enhances inside the microchannel. This trend is almost identical for all the configurations.

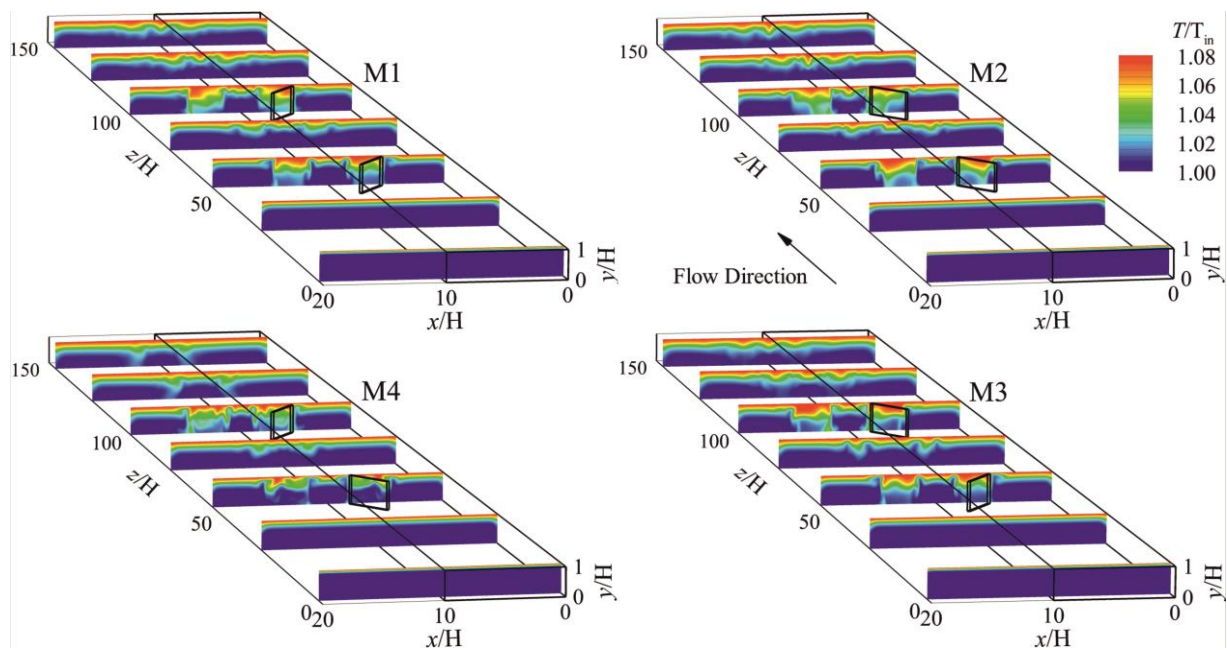

Figure 5- Contours of non-dimensional temperature ($T/T_{in}$) on different cross sections along the stream-wise direction for different configurations at $Re$=700.

Figure 6 indicates the effects of different configurations and Reynolds number on $Nu_m$. It demonstrates that the $Nu_m$ of the microchannels with LVGs is higher than that of the smooth microchannels in all ranges of the Reynolds numbers. Compared to M0, an augmentation in $Nu_m$ is seen between 10-14% for microchannels with LVGs at $Re$=700 and more enhancement is observed at higher Reynolds numbers. This is a sign that increasing the Reynolds number makes the vortices stronger, intensifies the fluid mixing and augments the swirl causing the $Nu_m$ to increase. One can notice that for M1 and M3 configurations the slopes of $Nu_m$ lines change rapidly when the $Re$ reaches about 700 while for M2 and M4 microchannels, it occurs when the $Re$ reaches about 500. It can be expressed that LVGs with an obtuse attack angle on the first row can induce stronger vortices in relation to one with acute attack angle. As the flow meets the first row of LVGs with higher velocity, the first row of LVGs plays more important role in the heat transfer enhancement. These generated vortices become stronger by increasing the Reynolds number and bring more heat transfer by decreasing the thermal boundary thickness, increasing flow disturbance, mixing and destabilization and enhancing the swirl in the microchannel.

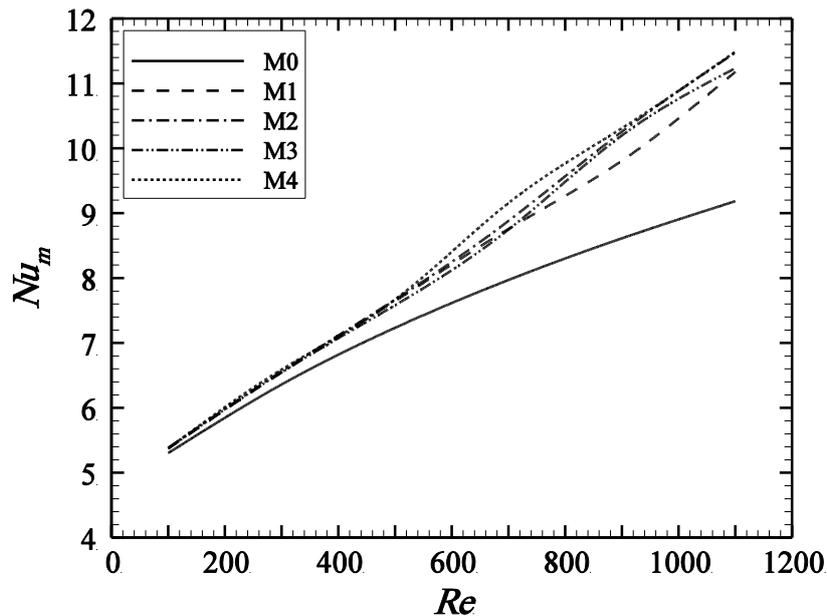

Figure 6- Effects of different configurations and Reynolds number on mean Nusselt number.

Despite having better thermal performance, microchannels equipped with LVGs cause higher pressure drop in the heat sink. Figure 7(a) depicts that using LVGs in microchannels leads to larger pressure penalty or apparent friction factor in the heat sink. This is due to the presence of secondary flow and complex interactions between the vortices and the microchannel walls. It is seen that all the configurations with LVGs have almost the identical apparent friction factor. It can be concluded that the friction resistance of microchannel walls is the most important factor influencing liquid flow resistance in the smooth microchannel; while, in addition of the frictional resistance of the channel walls, more pressure drag due to the local resistance of LVGs is brought. Figure 7(b) illustrates the required pumping power as a function of $Re$ for different configurations. In respect to the smooth microchannel, the maximum pumping power augmentation occurred at $Re$=1100 by about 30~39%. It is seen that higher pumping power is required at higher Reynolds numbers which can be attributed to augmented pressure loss across the microchannels and higher flow rate. The increase of $P_p$ is more obvious at high volume-averaged fluid velocity due to the direct relation between pressure drop and volume-averaged velocity of the working fluid.

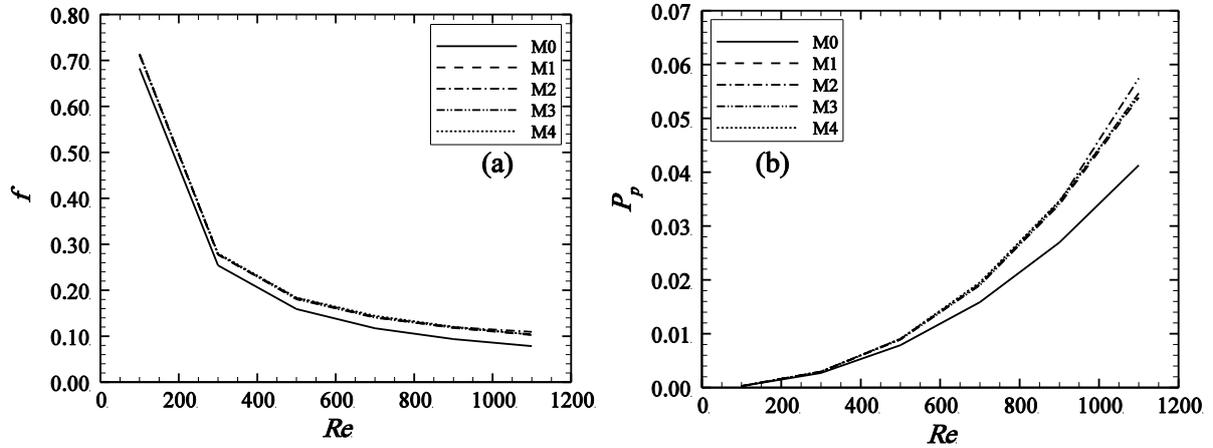

Figure 7- Variations of (a) apparent friction factor, (b) required pumping power, as a function of Reynolds number.

It is found that using LVGs in microchannels not only boosts the heat transfer performance, but also brings more pressure drop. In order to consider the overall efficiency [37-39] of the microchannels with LVGs the following parameters are defined.

$$\eta_{Nu} = Nu_m / (Nu_m)_s \qquad (13)$$

$$\eta_f = f / f_s \qquad (14)$$

$$\eta_T = \eta_{Nu} / \eta_f^{1/3} \qquad (15)$$

Where $\eta_{Nu}$, $\eta_f$ and $\eta_T$ are mean Nusselt number ratio, apparent friction factor ratio and overall efficiency, respectively. Moreover, subscript "s" indicates the smooth microchannel (case M0, see table 1). Figure 8 represents the summary of thermo-hydraulic performance of the microchannels with different configurations for various Reynolds numbers. It is seen that using LVGs in microchannels leads to a notable increase in overall efficiency and this increase will boost by increasing the Reynolds number. It is found that to achieve higher overall efficiency, it is better to use this technique under high flow rates. It is worth mentioning that among the all configurations with LVGs studied in the present study, M4 has the best average heat transfer performance and M2 has the highest average pressure drop. Moreover, M4 has the best average overall efficiency and after that M2, M3 and M1 are in the list, respectively.

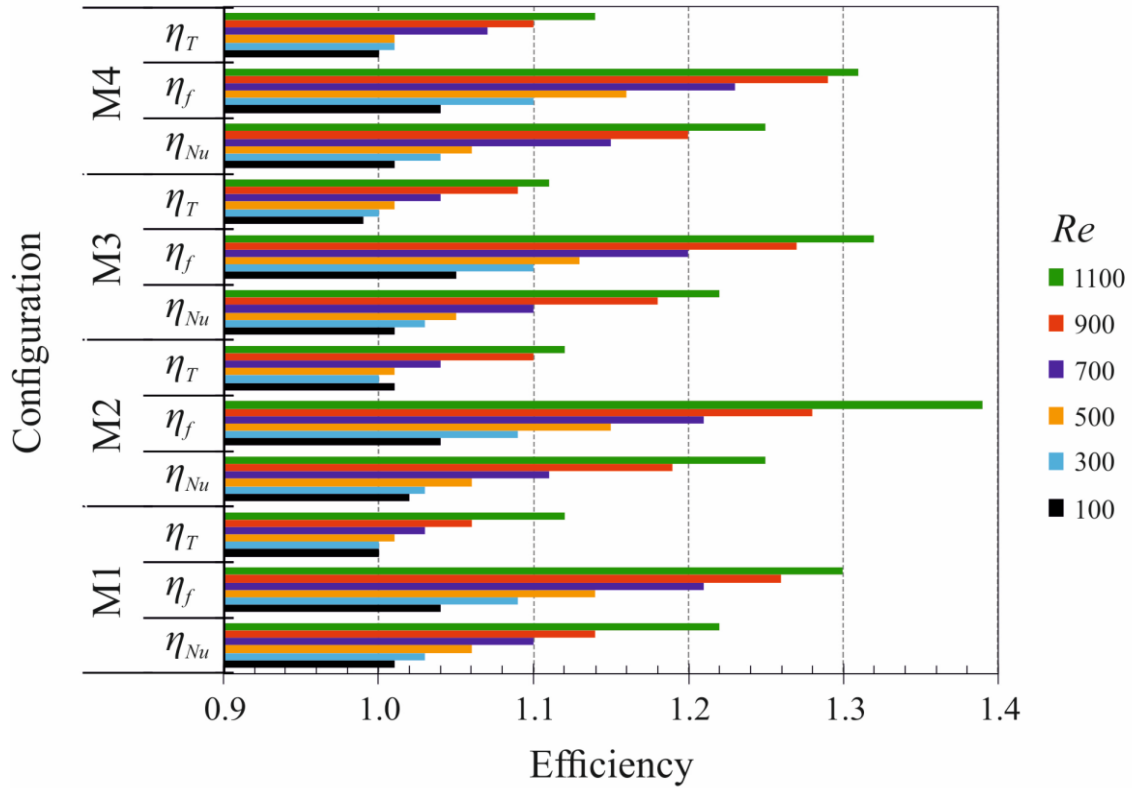

Figure 8- Effect of different configurations on mean Nusselt number ratio, apparent friction factor ratio and overall efficiency at different Reynolds numbers.

## 5. Conclusions

Three dimensional simulations were performed in a validated open-source computational fluid dynamics code, OpenFOAM, using finite volume approach. The effects of the Reynolds number and different geometrical configurations on the laminar single phase liquid flow and heat transfer in the microchannels with LVGs were investigated. The thermo-hydraulic performance of these types of microchannels was compared with the smooth microchannels and our numerical simulation obtained with laminar assumption agrees suitably with the experimental data. The following essential conclusions are acquired.

Using temperature-dependent thermo-physical properties results in higher mean Nusselt number compared to the constant thermo-physical properties and decreases the deviation from experimental results.

For the range of Reynolds numbers between 100 and 1100, a 2-25% increase in the mean Nusselt number was observed for microchannels with LVGs compared to smooth microchannels. This occurs due to better fluid mixing, reducing the thermal boundary-layer thickness and increase of heat transfer area. Furthermore, the friction factor was increased by 4-30% compared to smooth microchannels because of the local resistance of LVGs and presence of the secondary flow. Compared to the smooth rectangular microchannels, the application of the LVGs increases the required pumping power. Higher heat transfer and friction factor was observed at higher Reynolds numbers. Fluids with lower viscosity leads to better thermal performance in microchannels with LVGs. The effects of the winglets of the first row on flow pattern and heat transfer was more important inside the microchannels with LVGs.

The overall efficiency ($\eta_T$) of the microchannels equipped with LVGs is higher than 1 for all configurations, except one at *Re*=100, and whole range of the Reynolds numbers considered in the present study. Higher overall efficiency in microchannels with LVGs is attainable at high flow rates. Among the all configurations with LVGs studied in the present study, M4 has the best average overall efficiency and after that M2, M3 and M1 are in the list, respectively.

# References

1. Tuckerman, D.B. and R. Pease, *High-performance heat sinking for VLSI.* Electron Device Letters, IEEE, 1981. **2**(5): p. 126-129.

2. Morini, G.L., Single-phase convective heat transfer in microchannels: a review of experimental results. International Journal of Thermal Sciences, 2004. **43**(7): p. 631-651.

3. Agostini, B., et al., *State of the art of high heat flux cooling technologies.* Heat Transfer Engineering, 2007. **28**(4): p. 258-281.

4. Mohammed Adham, A., N. Mohd-Ghazali, and R. Ahmad, *Thermal and hydrodynamic analysis of microchannel heat sinks: A review.* Renewable and Sustainable Energy Reviews, 2013. **21**: p. 614-622.

5. Koyuncuoğlu, A., et al., Heat transfer and pressure drop experiments on CMOS compatible microchannel heat sinks for monolithic chip cooling applications. International Journal of Thermal Sciences, 2012. **56**: p. 77-85.

6. Harms, T.M., M.J. Kazmierczak, and F.M. Gerner, *Developing convective heat transfer in deep rectangular microchannels.* International Journal of Heat and Fluid Flow, 1999. **20**(2): p. 149-157.

7. Li, Z., W.-Q. Tao, and Y.-L. He, *A numerical study of laminar convective heat transfer in microchannel with non-circular cross-section.* International journal of thermal sciences, 2006. **45**(12): p. 1140-1148.

8. Webb, R.L. and N.-H. Kim, *Principles of enhanced heat transfer*. 2nd ed. 2005, Boca Raton: Taylor & Francis. xxii, 795 p.

9. Jacobi, A. and R. Shah, Heat transfer surface enhancement through the use of longitudinal vortices: a review of recent progress. Experimental Thermal and Fluid Science, 1995. **11**(3): p. 295-309.

10. Bergles, A.E., *Recent developments in enhanced heat transfer.* Heat and mass transfer, 2011. **47**(8): p. 1001-1008.

11. Johnson, T.R. and P.N. Joubert, The influence of vortex generators on the drag and heat transfer from a circular cylinder normal to an airstream. Journal of Heat Transfer, 1969. **91**: p. 91.

12. Ahmed, H., H.A. Mohammed, and M. Yusoff, *An overview on heat transfer augmentation using vortex generators and nanofluids: Approaches and applications.* Renewable and Sustainable Energy Reviews, 2012. **16**(8): p. 5951-5993.

13. Sohankar, A., *Heat transfer augmentation in a rectangular channel with a vee-shaped vortex generator.* International journal of heat and fluid flow, 2007. **28**(2): p. 306-317.

14. Sohankar, A. and L. Davidson, *Numerical study of heat and fluid flow in a plate-fin heat exchanger with vortex generators.* Turbulence Heat and Mass Transfer, 2003. **4**: p. 1155-1162.

15. Tian, L.-T., et al., Numerical study of fluid flow and heat transfer in a flat-plate channel with longitudinal vortex generators by applying field synergy principle analysis. International Communications in Heat and Mass Transfer, 2009. **36**(2): p. 111-120.

16. Fiebig, M., *Embedded vortices in internal flow: heat transfer and pressure loss enhancement.* International Journal of Heat and Fluid Flow, 1995. **16**(5): p. 376-388.

17. Biswas, G., H. Chattopadhyay, and A. Sinha, *Augmentation of heat transfer by creation of streamwise longitudinal vortices using vortex generators.* Heat Transfer Engineering, 2012. **33**(4-5): p. 406-424.

18. Fiebig, M., et al., *Heat transfer enhancement and drag by longitudinal vortex generators in channel flow.* Experimental Thermal and Fluid Science, 1991. **4**(1): p. 103-114.

19. Wu, J. and W. Tao, Effect of longitudinal vortex generator on heat transfer in rectangular channels. Applied Thermal Engineering, 2012. **37**: p. 67-72.

20. Khoshvaght-Aliabadi, M., F. Hormozi, and A. Zamzamian, Effects of geometrical parameters on performance of plate-fin heat exchanger: Vortex-generator as core surface and nanofluid as working media. Applied Thermal Engineering, 2014.

21. Ahmed, H.E. and M. Yusoff, Impact of Delta-Winglet Pair of Vortex Generators on the Thermal and Hydraulic Performance of a Triangular Channel Using Al2O3–Water Nanofluid. Journal of Heat Transfer, 2014. **136**(2): p. 021901.

22. Ahmed, H., H. Mohammed, and M. Yusoff, *Heat transfer enhancement of laminar nanofluids flow in a triangular duct using vortex generator.* Superlattices and Microstructures, 2012. **52**(3): p. 398-415.

23. Liu, C., et al., Experimental investigations on liquid flow and heat transfer in rectangular microchannel with longitudinal vortex generators. International Journal of Heat and Mass Transfer, 2011. **54**(13): p. 3069-3080.

24. Chen, C., et al., A study on fluid flow and heat transfer in rectangular microchannels with various longitudinal vortex generators. International Journal of Heat and Mass Transfer, 2014. **69**: p. 203-214.

25. Lan, J., Y. Xie, and D. Zhang, *Flow and Heat Transfer in Microchannels With Dimples and Protrusions.* Journal of heat transfer, 2012. **134**(2).

26. Mirzaee, H., et al., Heat Transfer Enhancement in Microchannels using an Elastic Vortex Generator. Journal of Enhanced Heat Transfer, 2012. **19**(3).

27. Hsiao, K.-Y., C.-Y. Wu, and Y.-T. Huang, *Fluid mixing in a microchannel with longitudinal vortex generators.* Chemical Engineering Journal, 2014. **235**: p. 27-36.

28. Kandlikar, S.G. and W.J. Grande, Evolution of Microchannel Flow Passages--Thermohydraulic Performance and Fabrication Technology. Heat transfer engineering, 2003. **24**(1): p. 3-17.

29. Ferrouillat, S., et al., Intensification of heat-transfer and mixing in multifunctional heat exchangers by artificially generated streamwise vorticity. Applied thermal engineering, 2006. **26**(16): p. 1820-1829.

30. C.J. Glassbrenner, G.A.S., Thermal conductivity of silicon and germanium from 3 K to the melting point. Physical Review 1964. **134** p. A1058–A1069.

31. A.S. Okhotin, A.S.P., V.V. Gorbachev, *Thermophysical Properties of Semiconductors.* 1972, Moscow: Atom Publicaion House.

32. W. Wagner, A.K., *Properties of Water and Steam, Springer-Verlag*. 1998, Berlin Heidelberg, Germany: Springer-Verlag.

33. Ma, J., et al., Experimental investigations on single-phase heat transfer enhancement with longitudinal vortices in narrow rectangular channel. Nuclear Engineering and Design, 2010. **240**(1): p. 92-102.

34. Seyf, H.R. and M. Feizbakhshi, Computational analysis of nanofluid effects on convective heat transfer enhancement of micro-pin-fin heat sinks. International Journal of Thermal Sciences, 2012. **58**: p. 168-179.

35. Bergman, T.L. and F.P. Incropera, *Fundamentals of heat and mass transfer*. 7th ed. 2011, Hoboken, NJ: Wiley. xxiii, 1048 p.

36. Kays, W.M., M.E. Crawford, and B. Weigand, *Convective heat and mass transfer*. 4th ed. McGraw-Hill series in mechanical engineering. 2005, Boston: McGraw-Hill Higher Education. xxx, 546 p.

37. Li, P., D. Zhang, and Y. Xie, Heat transfer and flow analysis of $Al_2O_3$–water nanofluids in microchannel with dimple and protrusion. International Journal of Heat and Mass Transfer, 2014. **73**: p. 456-467.

38. Nandi, T.K. and H. Chattopadhyay, *Numerical investigations of developing flow and heat transfer in raccoon type microchannels under inlet pulsation*. International Communications in Heat and Mass Transfer, 2014. **56**: p. 37-41.

39. Zhai, Y., et al., Heat transfer in the microchannels with fan-shaped reentrant cavities and different ribs based on field synergy principle and entropy generation analysis. International Journal of Heat and Mass Transfer, 2014. **68**: p. 224-233.

**Highlights**

- Temperature-depended thermo-physical properties is used to improve the numerical results.
- Thermo-hydraulic performance of the microchannels with LVGs is studied in details.
- Higher heat transfer enhancement is observed at higher Reynolds numbers.
- Microchannels with LVGs cause higher pressure loss in the device.
- The overall efficiency of the microchannels with LVGs is calculated.